\documentclass[preprint,...]{revtex4-2} 

\usepackage{tikz}
\usetikzlibrary{shapes,snakes,backgrounds,fit,decorations.pathreplacing}
\usepackage{bbm,times}
\usepackage{bm}
\usepackage{amsbsy}
\usepackage{amsthm}
\usepackage{amssymb}
\usepackage{amsfonts}
\usepackage{amsmath}
\usepackage[c]{esvect} 
\usepackage{dsfont} 
\usepackage{graphicx} 
\usepackage{epsfig}
\usepackage{epstopdf}
\usepackage{dsfont}
\usepackage{multibib}
\usepackage{color}
\usepackage{enumerate}
\usepackage{multirow}

\usepackage[colorlinks]{hyperref}
\makeatletter
\newcommand\org@hypertarget{}
\let\org@hypertarget\hypertarget
\renewcommand\hypertarget[2]{%
	\Hy@raisedlink{\org@hypertarget{#1}{}}#2%
}
\makeatother
\usepackage[figure,table]{hypcap}
\usepackage{MnSymbol}
\usepackage{enumerate}
\usepackage{float}
\hypersetup{
	bookmarksnumbered,   
	pdfstartview={FitH},
	citecolor={darkblue},
	linkcolor={darkred},
	urlcolor={darkblue},
	pdfpagemode={UseOutlines}}
\definecolor{darkgreen}{RGB}{50,190,50}
\definecolor{darkblue}{RGB}{0,0,190}
\definecolor{darkred}{RGB}{238,0,0}
\usepackage{soul}

\makeatletter
\renewcommand{\p@subsection}{}
\renewcommand{\p@subsubsection}{}
\makeatother

\begin{document}

\title{A mini review of NMR and MRI}

\author{Fatemeh Khashami
\\
\email{fatemeh.khashami@utsouthwestern.edu}
\\{\small Department of Internal Medicine, The University of Texas Southwestern Medical Center, Dallas, TX 75390\\
Department of Physics, The University of Texas at Dallas, Dallas, TX 75080
}
}

\date{\today}

\begin{abstract}
Nuclear magnetic resonance (NMR) and magnetic resonance imaging (MRI) are versatile tools with broad applications from physics and chemistry to geology and medical studies. In this mini-review, we consider the concepts of NMR and MRI technologies from their fundamental origins to applications in medical science.  We start from a quantum mechanical basis and consider the significant importance of NMR and MRI in clinical research. Furthermore, we briefly introduce different types of NMR systems. We also investigate some of the most important applications of MRI techniques to provide valuable methods for visualizing the inside of the body and soft tissues. 
\end{abstract}

\maketitle

\newpage

\section{Introduction}

Nuclear magnetic resonance (NMR) and magnetic resonance imaging (MRI) are two powerful technologies that help to understand biochemical information at molecular levels and investigate the inside of the human body. NMR is a non-destructive analytical technique that utilizes non-ionizing radio frequency (RF) waves for the chemical description of living and non-living objects. 
Particularly, MRI is widely operated in medical science for diagnostic imaging \cite{sakurai1995modern,griffiths2018introduction,feynman2010quantum,khashami2024fundamentals}.

NMR and MRI technologies are byproducts of our understanding of the atomic world and the discovery of quantum theory. In fact, it is impossible to comprehensively study the working principles of NMR and MRI without considering quantum physics. In particular, to explain the concepts behind these technologies, we need to understand spin in quantum physics \cite{maleki2020spin,maleki2021quantum1,maleki2021quantum2}.

The history of the discovery of quantum spin is one of the most exciting parts in the history of the modern science. This goes back to our understanding of matter in a more profound way, illuminated by the discovery of the elements of atoms. This pioneered by Rutherford's discovery of the 
atomic nucleus in $1911$ and the Bohr model of the atom in $1913$ \cite{hughes1990bohr,rutherford1932discussion}. Later, de Broglie's vision of the wave-particle duality challenged the classical physics viewpoint and revolutionized the perspective of quantum mechanics to a greater extent. This radical departure from the classical worldview makes the behavior of quantum systems counterintuitive, which still remains a subject of a heated debate \cite{maleki2023revisiting,maleki2021quantum,maleki2019stereographic}. The new perspective of quantum mechanics introduced a new vision of the physical world with manifesting probabilistic outcomes, superposition, and particle entanglement \cite{khashami2013entanglement,maleki2021natural}.

The foundation for quantum mechanics laid by Heisenberg's mechanics in $1925$ and Schrödinger's wave equation in $1926$. In the same year, Uhlenbeck and Goudsmit discovered electron spin and its intrinsic angular momentum properties. 
Regarding electron spin and its properties in a quantum system, Paul Dirac revised the hydrogen atom's spectrum in $1928$. Israel Isaac Rabi initiated the principle of NMR technology in $1938$, known as Rabi's discovery. Furthermore,  in $1946$, Felix Bloch and Edward Purcell developed the first NMR machines. The foundation for NMR is related to the understanding of the Zeeman effect, illustrating how the interaction of nuclear spins with external magnetic fields played a crucial role in developing quantum mechanics \cite{bloch1946packard,friston1994analysis,roberts1977basic}.

The Bloch equation explains the dynamics of the NMR system and describes the spin relaxation dynamics in a quantum spin system. Later, the concept of the NMR signal is described by the Fourier transform (FT) method, corresponding to the free induction decay (FID) signal, to convert the time domain NMR signal to the frequency domain signal. Additionally, in $1991$, Richard R. Ernst followed the FT-NMR method and developed the FT-NMR spectroscopy \cite{feynman2010quantum,butz2006fourier,rabi1938new}. 

It is important to highlight that NMR was the first generation of Rabi's discovery, whereas MRI represents the second generation of this phenomenon. These developed technologies provide practical tools for understanding and describing various medical issues, opening the new way to medical imaging as a diagnostic device for investigating cancerous and normal tissues \cite{graham1985anaerobic,abragam1983principles,solomon1955relaxation}.

In this review, we focus on the quantum mechanical foundations of NMR and MRI and attempt to design an understandable foundation for the quantum mechanical framework of these technologies. In order words, this review aims to bridge the gap between the technological application of quantum mechanics and the fundamental principles based on which these technologies are built. Learning the basic frameworks behind the technologies is beneficial for many reasons \cite{khashami2024fundamentals,khashami2021tracking}.

\section{The Zeeman Effect of Nuclear Spins}

\begin{figure}
\centering
	\includegraphics[width=\linewidth]{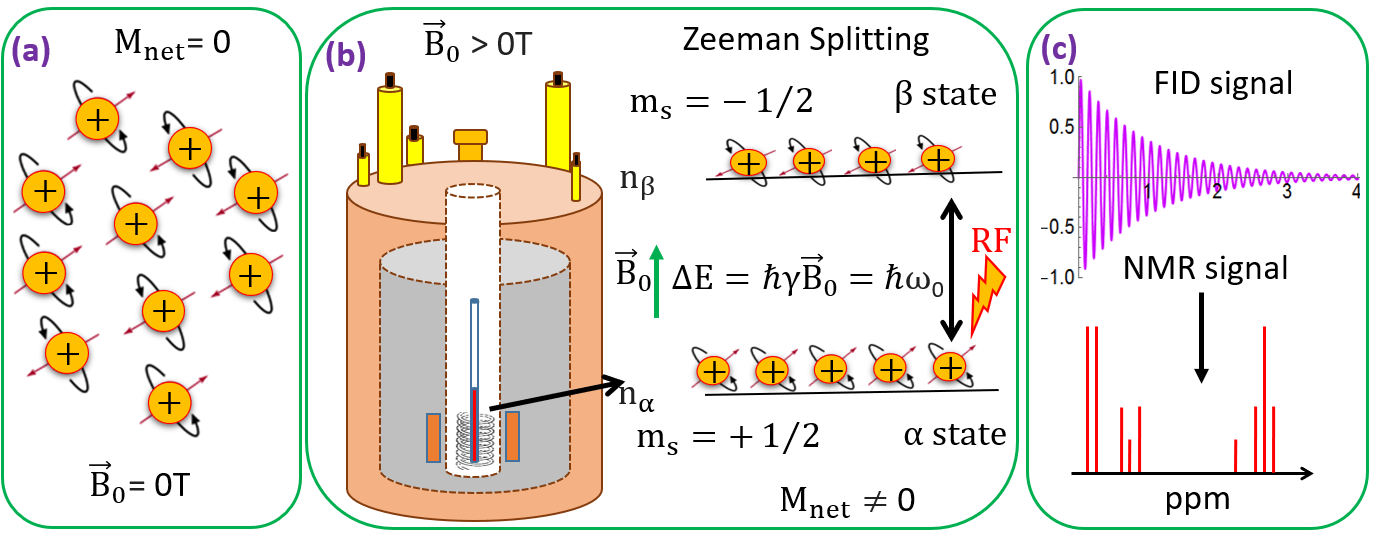}
	\caption{(a) In the absence of the magnetic field $\vec{B_0}$, ${}^{1}H$ atoms lay in random orientations, and their nuclei occupy a single energy level. (b) Due to Zeeman splitting, when the strong magnetic field is applied, some nuclei align parallel to $\vec{B_0}$ ($\alpha$ state) and others anti-parallel $\vec{B_0}$ ($\beta$ state). More of the nuclei occupy the lower energy level. After applying an RF pulse, some nuclei in the lower state move to the higher. (c) When the RF pulse is switched off, the nuclei return to equilibrium and emit a free induction decay (FID) signal. By the Fourier transformation (FT) method, the FID signal converts to the NMR spectrum 
    }
	\label{zeeman}
\end{figure}

To start the physics behind NMR and MRI technologies, we can consider a spin system's behavior in the absence and presence of a magnetic field, $\vec{B_0}$. Spins with magnetic dipole moments $\mu_s$ are oriented in random directions that are out of phase in general in the absence of a magnetic field. Moreover, their net magnetic moment is zero \cite{abragam1983principles,mansfield1977multi,khashami2021tracking}, as we presented in Fig. \ref{zeeman}(a). 
After applying an external magnetic field, $\vec{B_0}$, the nuclei precess in phase, which can be along with or against the magnetic field relying on the spin of the nuclei, which is shown in Fig. \ref{zeeman}(b) \cite{emsley2013high,maleki2023revisiting}. 

The magnetic field builds a torque on $\mu_s$ of the nuclei and orients it along the $z$-axis. The nuclear magnetic moment of the nucleus spin aligns with or against $\vec{B_0}$, which has ($2I+1$) a number of states, which results in Zeeman splitting. This splitting creates ($2I+1$) energy levels that are expressed by the spin quantum number, $m_s$, which has the magnitude $1/2$ and direction $+$ or $-$ along the $z$-axis \cite{davies1991book,ames2010theory,lancaster2014quantum,maleki2016generation,khashami2024fundamentals}.

From the Zeeman effect, the NMR signal depends on the energy difference between the two-level states of the nucleus. The lower energy level is known as the $\alpha$ state, with $m_s = +1/2$, for which the nucleus is aligned with the magnetic field, and the higher energy level is known as the $\beta$ state, with $m_s$ = $-1/2$, for which the proton is aligned against $\vec{B_0}$.  The lower energy level often has slightly more occupants than the higher energy level, in general, \cite{ames2010theory,purcell1946resonance,bloch1946packard,rabi1938new}.

The spin angular momentum in $z$-axis is $S_z= m_s \hbar = \pm \hbar/2$, and the total magnitude spin angular momentum vector for spin quantum number $1/2$ is $|\vec{S}|=  \sqrt{s(s+1)}\hbar = \sqrt{3/4}\hbar$, where $\hbar$ is Planck constant, which is equal to $1.054 \times 10^{-34}$ \text{J}.\text{S}.  The magnetic dipole moment of nuclei, $\mu_{z}$, with $m_s=\pm1/2$,  is defined as $\mu_z= m_s \hbar \gamma = \pm1/2 \hbar \gamma\label{mu}$, where $ \gamma $ is gyromagnetic ratio with units MHz/Tesla, and it is a constant value for each nucleus and is related to the magnetic moment $\mu_s$, and the spin number $I$ for a specific nucleus is expressed as
\begin{equation}
\gamma=\frac{2\pi\mu_z}{hI},
\end{equation}
where $\mathrm{\gamma}$ can be positive or negative \cite{damadian1980field,kitaev2002classical,de1970reinterpretation,khashami2024fundamentals}.   

Due to the Zeeman splitting, the interaction energy of a spin state with a magnetic field along the $z$-axis is directly proportional to the magnetic field strength and is expressed as 
\begin{equation}
E= - \vec{{\mu}}.\vec{{B_0}}.
\end{equation}
The energy along the $z$-axis is  given by $E= - \vec{{\mu}_z}.\vec{{B_0}}$. Therefore, the energy difference between the two upper and lower energy levels is obtained as
\begin{equation}
\Delta E = E_{\beta}-E_{{\alpha}} = \hbar\gamma \vec{\mathrm{B_0}}=\hbar \omega_0.
\label{energy}
\end{equation}

In this relation, the nucleus spin precess at a specific frequency $\omega_0$, known as the Larmor frequency, has the unit rad/s. As we can see from Eq. (\ref{energy}), the Larmor frequency, which
is the precession frequency of the spins around the axis of the magnetic field, is $\omega_0=\gamma \vec{{B_0}}$ \cite{jackson1999classical,khashami2024fundamentals}. Using resonant RF pulses, we can excite nuclei in the lower energy state  \cite{lauterbur1973image,bushong2013magnetic,hashemi2010basic}. An NMR spectrum usually uses RF pulses in frequency between 60-900 MHz \cite{abragam1983principles,damadian1980field,mansfield1977multi}. When the RF pulse is turned off, the nuclei precess around  $\vec{{B_0}}$ \cite{hashemi2010basic,heisenberg1989encounters}.

The Zeeman splitting leads to a net magnetization of the sample, which means that once the magnetic field is turned off, the nuclei flip down, and the magnetization vector approaches equilibrium. Thus, the rotating magnetization vector generates a current in a receiver coil. This current makes the detection of the NMR signal possible.
The oscillating signal presents waves with decreasing magnitude as the nucleus re-aligns with the magnetic field. This wave is called a free induction decay (FID) signal. Then, the Fourier transformation (FT) method converts the FID signal to the frequency domain, which appears as peaks along the $x$-axis of the NMR spectrum. This is called chemical shift and is given in units ppm [see Fig. \ref{zeeman}(c)] \cite{solomon1955relaxation,hashemi2010basic}. We will explain the chemical shift properties later in this review.

\section{The Boltzmann Distribution of Nuclear Spins}

The Boltzmann distribution describes the number of nuclei in each spin state. At room temperature, the number of spins in the lower energy level (n$_{\alpha}$) is greater than the number in the higher energy level (n$_{\beta}$). However, the energy separation between these two states is relatively small. Therefore, the number of nuclei at each spin state depends upon the temperature of the sample \cite{bushong2013magnetic,solomon1955relaxation,maleki2015entanglement,khashami2024fundamentals}. 

The Boltzmann distribution shows that the NMR signal intensity is proportional to the applied magnetic field and inverse to the temperature. The weak magnetic moments of nuclear spins cause small differences in the nuclear spin populations \cite{levitt1982broadband,schwabl2013statistical,maleki2019quantum}. In other words, the NMR signal is proportional to the nuclear polarization $\mathrm{P}$, governed by the Boltzmann distribution of nuclear spins on the Zeeman energy levels. In thermal equilibrium, for a spin$-1/2 $ system, it is determined by a thermal population difference determined by the Boltzmann factor as
\begin{equation}
P= \tanh \frac{\gamma\hbar B_0}{2k_BT},
\end{equation}
where $P$ is the spin polarization, $T$ is the temperature, and $k_B$ is the Boltzmann constant.
The relative number of upper and lower population is given by $
n_{\alpha}+n_{{\beta}}=n_{0}$, where $n_{0}$ is the number of nuclei per unit volume \cite{stoneham1969shapes,levitt2013spin,davies1991book}. 

At thermodynamic equilibrium, the ratio between the populations of the two levels can be written as
\begin{equation}
\frac{n_\alpha}{n_{\beta}}= 
\exp\left({\dfrac{-\Delta E}{k_BT}}\right)=\exp\left({\dfrac{-\hbar \omega_0}{k_BT}}\right)\approx 1-\dfrac{\hbar \omega_0}{k_BT}.
\label{kel}
\end{equation}

From Eq. (\ref{kel}) we have \cite{khashami2024fundamentals}
\begin{align}
\nonumber
n_{\beta}&=\frac{n_{0}}{1+e^{\beta\hbar {\omega}_0}} \cong \frac{n_{0}}{2+\beta\hbar {\omega}_0}=\frac{n_{0}}{2} \frac{1}{1+{\beta\hbar {\omega}_0 / 2}} \cong \frac{n_{0}}{2}(1-\frac{\beta\hbar {\omega}_0}{ 2}),
\\
n_{\alpha}&=\frac{n_{0}}{1+e^{-\beta\hbar {\omega}_0}} \cong \frac{n_{0}}{2-\beta\hbar {\omega}_0}=\frac{n_{0}}{2} \frac{1}{1-{\beta\hbar {\omega}_0 / 2}} \cong \frac{n_{0}}{2}(1+\frac{\beta\hbar {\omega}_0}{ 2}).
 \label{lower}
\end{align}

Moreover, the magnetization can be calculated for the spin $1/2$ system in terms of the ${T}$ and $\vec{{B_0}}$. The total magnetization is expressed as $M_{0}=n_{\alpha}\mu_{z}^{(\alpha)}+n_{\beta}\mu_{z}^{(\beta)}$, that is the sum of all the spins in the sample known as Bulk magnetization. At equilibrium, the magnetization aligned with $\vec{{B_0}}$ is given by Curie’s law of spin $1/2$ nuclei as
\begin{equation}
M_{0}=\left(n_{\alpha}-n_{\beta}\right) \frac{\gamma \hbar}{2} \cong \frac{\gamma \hbar}{2} n_0 \frac{\hbar \omega_{0}}{2k_{B} T}=\frac{n_{0} \gamma^{2} \hbar^{2}}{4 k_{B} T}B_{0}= \chi_0 B_0,
\label{Mag1}    
\end{equation}
where $\chi_0$ is the static nuclear susceptibility of a sample, which shows how the system becomes magnetized in $\vec{{B_0}}$, and it is inversely proportional to the ${T}$. 

\section{Chemical Shift}

\begin{figure}
\centering
	\includegraphics[width=\linewidth]{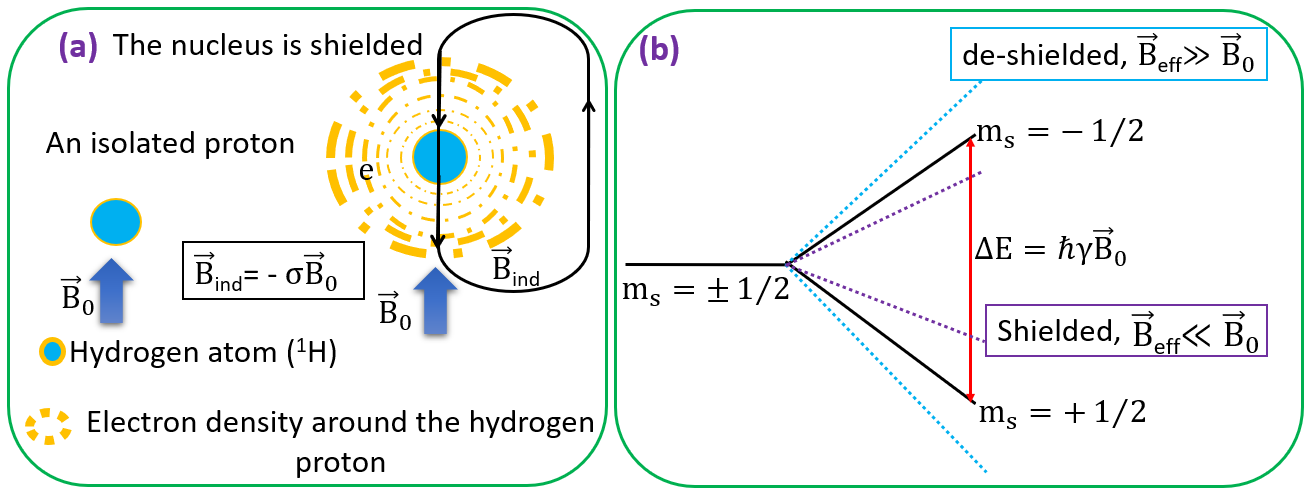}
	\caption{ (a) In  Hydrogen nuclei, electrons change the effective magnetic field ($\vec{B}_{\text {eff}}$) experienced by atoms. The external field ($\vec{B}_0$) induces currents around the nucleus. The induced magnetic field ($\vec{B}_{\text {ind }}$) is opposite in the direction to  $\vec{B}_0$, and the electrons shield the nucleus. (b) For spin quantum number $1/2$, there is ($2I+1$) degeneracy, the energy difference between the two states is $\Delta E = \hbar\gamma \vec{\mathrm{B_0}}$. When $\vec{B}_{\text {eff}}$ is less than  $\vec{B}_0$,  the nucleus becomes shielded, and when $\vec{B}_{\text {eff}}$ is more than $\vec{B}_0$, it becomes de-shielded
    }
	\label{fig:9}
\end{figure}

The chemical shift can determine molecules' structure and functional groups in the NMR spectrum. Since the chemical shifts represent protons, they must absorb different amounts of energy to attain many peaks due to magnetic shielding. All protons are surrounded by electrons that shield them from  $\vec{{B_0}}$, and the more circulating electrons, the more shielding effect occurs [see Fig. \ref{fig:9}(a)] \cite{bakhmutov2004and,rabi1938new,sakurai1995modern}. The amount of shielding is directly associated with the electron density around  ${}^{1}H$ nucleus. The higher shielded electrons need higher energy for the resonance. 

The effective magnetic field, $\vec{B}_{\text {eff}}$, experienced by a nucleus is the sum of  $\vec{{B_0}}$ and the induced magnetic field, $\vec{B}_{\text {ind }}$, which is generally less than  $\vec{{B_0}}$, and can be expressed as
\begin{equation}
\vec{B}_{\text {eff}} = \vec{B}_0 + \vec{B}_{\text {ind }}= \vec{B}_0(1-\sigma),
\end{equation}
where $\sigma$ is the shielding constant.
According to Lenz's Law,  $\vec{{B_0}}$ that influence the electron density of the proton of  ${}^{1}H$ atom that opposes  $\vec{{B}}_{\text {ind }}$ is given by $\vec{{B}}_{\text {ind }}=-\sigma \vec{B_0}$.
The nucleus in various local environments experiences slightly different frequencies from $\omega_0$, which is called effective Larmor frequency $\omega_{eff}$ \cite{bakhmutov2004and,lancaster2014quantum,shifrin1998new}. The shielding constant for a particular nucleus in a particular environment is given by
\begin{equation}
\omega_{eff}=\gamma \vec{B}_{\text {eff}}=\gamma \vec{B}_0(1-\sigma).
\label{sheld1}
\end{equation}

From Fig. \ref{fig:9}(b), when  $\vec{B}_{\text {eff}}$
at the position of the nucleus is less than $\vec{{B_0}}$, the nucleus gets shielded. Otherwise, the nucleus is considered de-shielded.  In the shielding, the electron density circulates about the direction of  $\vec{{B_0}}$ and
opposes the external magnetic field, reducing  $\vec{B}_{\text {eff}}$. As a result of the electron shielding, the nucleus experiences different frequencies, commonly known as the chemical shift, denoted by $\delta $, and expressed in parts-per-million (ppm).

The chemical shift of sample, ${\delta_{s}}$ with the resonance Larmor frequency of the sample ${{\omega_{s}}}$, is independent of $\vec{{B_0}}$ and is dependent on the environment of the nucleus. By considering the resonance Larmor frequency of an internal reference, ${{\omega_{ref}}}$, the chemical shift of the sample in ppm can be attained as 
\begin{equation}
{\delta_s}= 10^6 \times \dfrac{\omega_s - \omega_{ref}}{\omega_{ref}}.
\end{equation}

\section{The Bloch equations description} \label{Relaxation}

The signal measured in an NMR sample is due to the existence of net magnetization, $\vec{M}$, which was first proposed in $1946$ by Felix Bloch \cite{bloch1946packard,posener1959shape,purcell1946resonance}. Central to the NMR technology is the concept of the relaxation of the net magnetization. The relaxation is due to spin-lattice and spin-spin relaxations in the system. Therefore, in this process, the spins decohere and relax to their ground states \cite{maleki2021perfect,maleki2018recovery,maleki2017entanglement}

\begin{figure}
\centering
	\includegraphics[width=\linewidth]{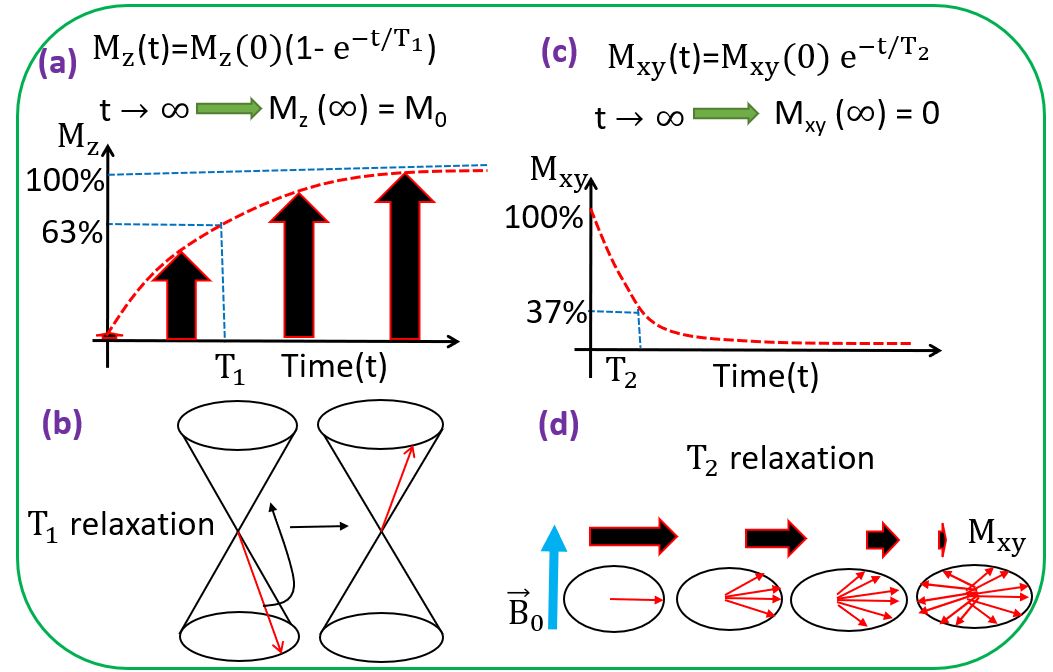}
	\caption{(a)-(b) When an RF pulse is applied, some spins flip to the higher energy state, decreasing $M_{z}$. If the RF pulse is turned off, $M_{z}$ increases gradually and reaches the initial magnetization $M_{0}$ at infinity. When the magnetization reaches to $\sim63\% $ of $M_{z}$ it records $T_1$. (c)-(d) Once the RF pulse is switched off, the magnetization vector in $x-y$ plane, $M_{xy}$ returns to its initial equilibrium (its maximum value) condition and starts processing in the $x-y$ plane at different Larmor frequencies, recording $T_2$ relaxation time at $\sim37\% $ of $M_{xy}$ 
    }
	\label{relaxation}
\end{figure}

Spin-lattice relaxation or longitudinal relaxation is the process by which the net magnetization returns to the $z$-axis, known as the longitudinal magnetization, $M_z$. In this scenario, the net magnetization is aligned with the external magnetic field \cite{stoneham1969shapes,vathyam1996homogeneous}. This process is due to energy exchange between the spin system and neighboring molecules. By switching off the RF pulses, $\sim63\% $ of the magnetization returns to thermal equilibrium \cite{levy1975experimental,sepponen1985method}. The spin-lattice relaxation rate is governed by a constant value, $T_1$, an important quantity in NMR spectroscopy. $T_1$ determines how quickly the spin of the nucleus becomes parallel to the magnetic field [see Fig. \ref{relaxation}(a)-(b)]
\cite{shrivastava1983theory,bakhmutov2004and}.

On the other hand, spin-spin relaxation or transverse relaxation is a process due to interaction between the spin system and their magnetic field on the $x-y$ plane, known as transverse magnetization, $M_{xy}$. When an RF pulse is switched off, $M_{xy}$ decays to zero with the time constant $T_2$, where only $\sim37\% $ of original $M_{xy}$ is present [see Fig. \ref{relaxation}(c)-(d)] \cite{solomon1955relaxation,vathyam1996homogeneous,bushong2013magnetic,orbach1961spin}. 

In the relaxation process, the magnetization returns to equilibrium along the $z$-axis and in the $x-y$ plane at different rates. According to the Bloch equation, the dynamics of the net magnetization is given by
\begin{equation}
    \frac{d \overrightarrow{M}}{d t}=\overrightarrow{M} \times \gamma \overrightarrow{B}-\frac{M_{x} \hat{i}+M_{y} \hat{j}}{T_{2}}-\frac{\left(M_{z}-M_{0}\right) \hat{k}}{T_{1}},
\end{equation}
where $M_x$, $M_y$, and $M_z$ are the components of magnetization $\vec{M}$ in the $x, y$ and $z$ directions, and $M_0$ is the magnetisation at the steady state. Also, $T_1$ and $T_2$ are the spin-lattice and spin-spin relaxation times, respectively. Considering each of the Bloch components in magnetic field ($B_x$, $B_y$, $B_z$), we attain
\begin{align}
\frac{d M_{x}(t)}{d t} &=\gamma\left(M_{y}(t) B_{z}(t)-M_{z}(t) B_{y}(t)\right)-\frac{M_{x}(t)}{T_{2}}, \\
\frac{d M_{y}(t)}{d t} &=\gamma\left(M_{z}(t) B_{x}(t)-M_{x}(t) B_{z}(t)\right)-\frac{M_{y}(t)}{T_{2}}, \\
\frac{d M_{z}(t)}{d t} &=\gamma\left(M_{x}(t) B_{y}(t)-M_{y}(t) B_{x}(t)\right)-\frac{M_{z}(t)-M_{0}}{T_{1}}.
\label{Mag}
\end{align}

For $B_{x}=0$, $B_{y}=0$, and $B_{z}=B_{0}$, the rate of change of magnetization from the Bloch equation becomes
 \begin{align}
&\frac{d M_{x}(t)}{dt} =\omega_0 M_{y}(t)-\frac{M_{x}(t)}{T_{2}}, \\
&\frac{d M_{y}(t)}{dt} = -\omega_0 M_{x}(t)-\frac{M_{y}(t)}{T_{2}},\\
&\frac{d M_{z}(t)}{dt} =\frac{M_{0}-M_{z}(t)}{T_{1}}.
\label{z}
 \end{align}

Eq. (\ref{z}) is known as the Bloch equation for $z$-magnetization, which characterizes the longitudinal component of the magnetization \cite{tipler2003modern,jackson1999classical,shrivastava1983theory}. The rate of change of $z$-magnetization with time is inversely proportional to $T_1$. It is directly proportional to the difference between the equilibrium value of magnetization, $M_{0}$, and $z$-magnetization at time $t$, $M_{z}(t)$ [see Fig. \ref{relaxation}(a)]. The simplified solution for $M_{z}(t)$ reads 
\begin{equation}
    M_{z}(t)=M_{z}(0) \exp\left(-\frac{t}{T_{1}}\right)+M_{0}\left(1-\exp\left(-\frac{t}{T_{1}}\right)\right).
\end{equation}
At time $t=0$ with $M_{z}(0)=0$, this expression simplifies to 
\begin{equation}
    M_{z}(t)= M_{0}\left(1-\exp\left(-\frac{t}{T_{1}}\right)\right),
    \label{eq:Mz}
\end{equation}
which indicates that $M_{z}(t)$ increases with time, and exponentially approaches the equilibrium value $M_{0}$. Equation (\ref{eq:Mz}) shows that the magnetization enhances to $(1- e^{-1})\sim63\%$ of its maximum value in time constant $T_1$. When $t<<T_1$, the equation can be further reduced as $M_{z}(t)=  (t/T_1)M_{0}$. Moreover, the longitudinal magnetization at $t\rightarrow{\infty}$ reaches to its equilibrium value $M_{z}(\infty)= M_{0}$.

\begin{figure}[hh]
\centering
	\includegraphics[width=\linewidth]{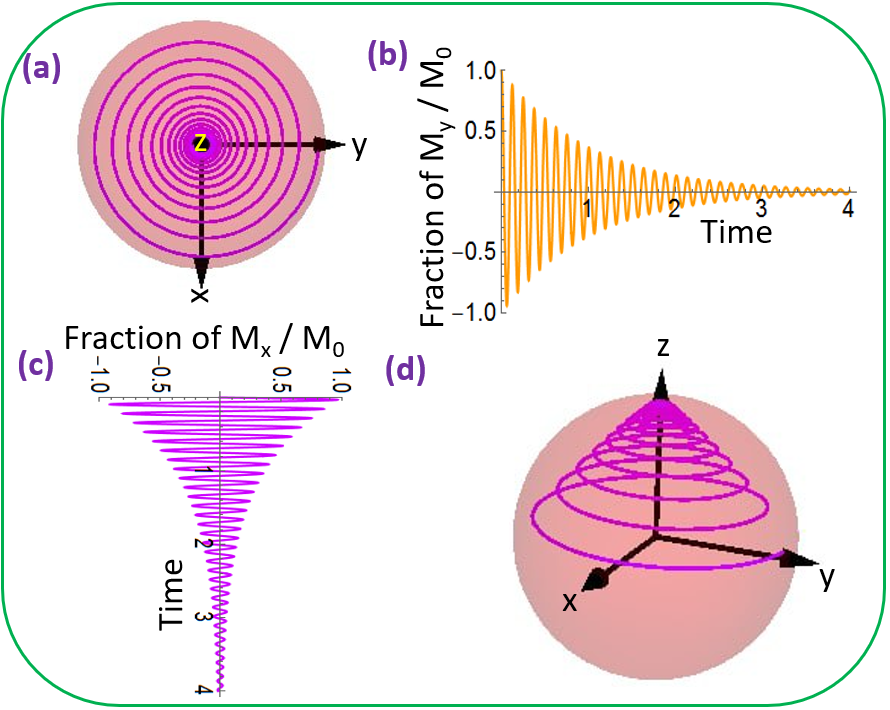}
	\caption{When an RF pulse is applied, the FID signal is collected. The FT method converts the FID signal to the NMR spectrum: (a) Transverse magnetization in the $x-y$ plane, where the external magnetic field is in the $z$-axis. (b)-(c) Present the FID signal along the $y$-axis and $x$-axis, respectively, where their combination provides (a). (d) The NMR signal is in three dimensions
    }
	\label{fig:2}
\end{figure}
The magnetization in the transverse plane is a complex value as $M_{x y}(t)=M_{y}(t)+i M_{x}(t)$. The solutions for $M_{xy}(t)$ can be found by solving the differential equations given by the Bloch equation as
\begin{align}
\nonumber
    M_{x y}(t)&=M_{x y}(0)\exp \left(-i\omega_0 t\right) \exp\left(-\frac{t}{T_{2}}\right) 
    \\
    &
    \nonumber
    =M_0 \exp \left(-i\omega_0 t\right) \exp \left(-\frac{t}{T_{2}}\right)
     \\
    &
    = M_0 (\cos (\omega_0 t) - i \sin (\omega_0 t))\exp \left(-\frac{t}{T_{2}}\right).
\label{FID}  
\end{align}

We also mention that $M_{xy}$ at $t=0$ is defined by $M_{xy}(0)=$ $M_0$. $\operatorname{Re}(M_{x y}(t))$ and $\operatorname{Im}(M_{x y}(t))$ are functions that return the real and imaginary parts of complex number $M_{x y}$ \cite{mcintyre2002spin,wang1986pictorial}. Therefore
\begin{align}
M_{x}(t)=\operatorname{Im}\left(M_{x y}(t)\right)=- M_0 \sin \left(\omega_{0} t\right)\exp \left(-\frac{t}{T_{2}}\right), \\
M_{y}(t)=\operatorname{Re}\left(M_{x y}(t)\right)=M_0 \cos \left(\omega_{0} t\right)\exp \left(-\frac{t}{T_{2}}\right).    
\end{align}

Once $M_{xy}$ exists, magnetization quickly starts vanishing by
\begin{equation}
M_{x y}(t)=M_0 \exp \left(-\frac{t}{T_{2}}\right).
\label{exp}   
\end{equation}
Where $M_{xy}$ vanishs at $t \rightarrow{\infty}$, as $M_{x}(\infty)= M_{y}(\infty)= 0$ [see Fig. \ref{relaxation}(b)]. Consequently, the characteristic time $T_2$ refers to the time it takes for the magnetization to drop to ($1/e$) of its original value in the presence of the homogeneity field. 

A schematic of the dynamics of the elements of the Bloch equation is presented in Fig. \ref{fig:2}. This figure provides a geometric visualization of the relaxation process of the NMR magnetization. Accordingly, the system's dynamics in the $x-y$ plane are given in Fig. \ref{fig:2}(a). As can readily be seen, the magnetization starting from the $y$-axis direction processes along the  $z$-axis and decays to the center of the disc in the $x-y$ plane. This process from the point of view of the $y$- and $x$-axis can be understood according to 
Fig. \ref{fig:2}(b) and (c), respectively. Including the dynamics in the $z$-axis, the entire process can be visualized via Fig. \ref{fig:2}(d).
The comparative dynamics of the $y$- and $x$-axis for the initial magnetization in the $y$-axis direction is depicted in  Fig. \ref{fig:2}(e).

Equation (\ref{exp}) is known as the FID signal, which is composed of a decaying oscillation sine and cosine functions [see Fig. \ref{fig:2}(a)-(d)]. By the FT method, F$(\omega)$, the FID signal converts to the frequency domain and appears as several peaks on the  NMR spectrum. The FT method can be expressed as $F(\omega)=\int_{-\infty}^{+\infty} f(t) \exp (i \omega t) d t$.   F$(\omega)$ is a complex function that can be divided into real and imaginary parts as
\begin{equation}
\operatorname{Re}(F(\omega))=\int_{-\infty}^{+\infty}f(t)\cos (\omega t)dt,\quad
\operatorname{Im}(F(\omega))=\int_{-\infty}^{+\infty}f(t)\sin (\omega t)dt. 
\end{equation}

\section{Different types of NMR spectroscopy}

\begin{table}[ht]
\caption{Different types of nucleus with spin quantum number $1/2$ and their properties}
\centering 
\begin{tabular}{|c|c|c|c|}
\hline     \text{Nucleus} & $\gamma (T^{-1} s^{-1})$ &  \text{Nat. Abd (\%)} \\[0.5ex]
\hline     
${ }^{1}H$  &  2.6752  $\times$ 10$^8$ &  100.0  \\
${ }^{13}C$ & 6.7283 $\times$ 10$^7$ &  1.11  \\
${ }^{19}F$ & 2.517  $\times$ 10$^8$ & 100.0  \\
${ }^{15}N$ & -2.713 $\times$ 10$^7$  & 0.37 \\
${ }^{31}P$ & 1.0841 $\times$ 10$^8$ & 100.0  \\ [1ex]
\hline
\end{tabular}
\label{nat1}
\end{table}

The NMR spectroscopy is a versatile technique that is utilized to study a variety of atomic isotopes with nuclear spin $1/2$ such as ${{}^{1}}H$, ${{}^{13}}C$, ${{}^{31}}P$, ${{}^{15}}N$ and ${{}^{19}}F$. In Table \ref{nat1}, we list different types of the nucleus with the spin quantum number $1/2$ and present their gyromagnetic ratio $\gamma$ and natural abundance (Nat. Abd ($\%$)) \cite{golman2006metabolic,filler2009history,roberts1977basic}. 
In this section, we will consider NMR spectroscopy using the atomic isotopes in Table \ref{nat1}.

\subsection{Proton Nuclear Magnetic Resonance Spectroscopy}

\begin{figure}[ht]
\centering
	\includegraphics[width=\linewidth]{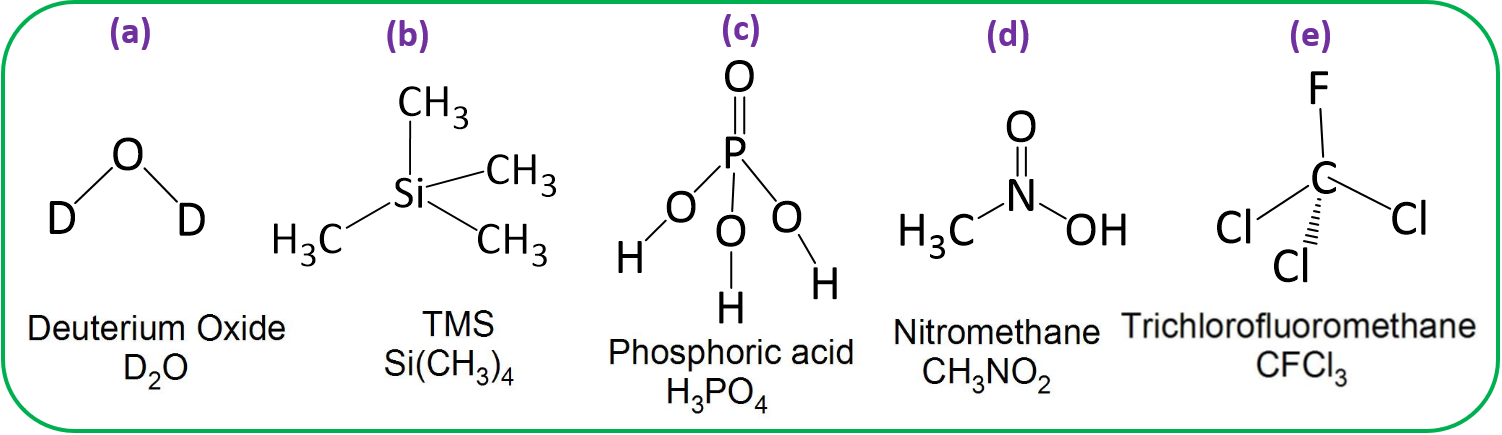}
	\caption{Scheme of the internal standard reference for different types of NMR: (a) Deuterium oxide ($D_2O$) for ${{}^{1}}H$-NMR, (b) TMS ($Si(CH_3)_4$) for ${{}^{13}}C$-NMR, (c) Phosphoric acid ($H_3PO_4$) for ${{}^{31}}P$-NMR, (d) Nitromethane ($CH_3NO_2$) for ${{}^{15}}N$-NMR (e) Trichlorofluoromethane ($CFCl_3$) for ${{}^{19}}F$-NMR
	}
	\label{reff}
\end{figure}

\begin{figure}[ht]
\centering
	\includegraphics[width=\linewidth]{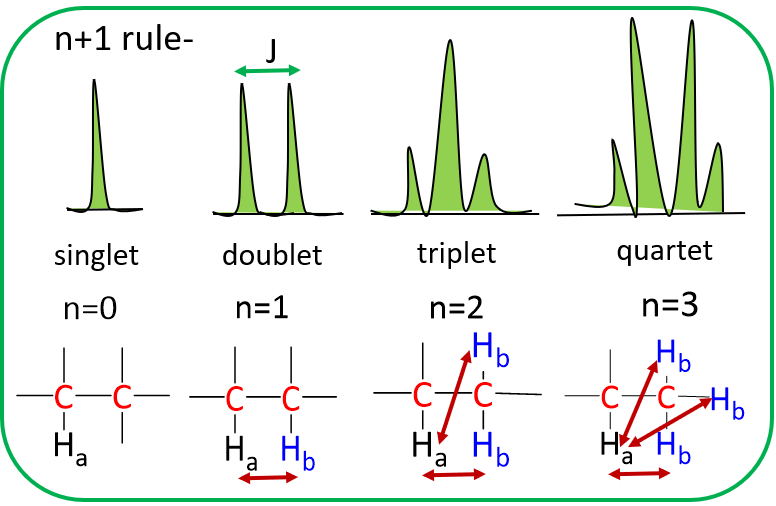}
	\caption{Splitting patterns in NMR spectroscopy: By $n+1$ rule, the diversity of the signal is calculated. Singlet signals are not coupled to any protons; doublet signals are coupled to one proton; triplet signals are coupled to two protons; quartet signals are coupled to three protons. The coupling constant, J, is the distance between the peaks in a splitting pattern. Hydrogen nuclei $H_a$, $H_b$  
	}
	\label{splitting}
\end{figure}

Proton NMR or ${ }^{1} {H}$-NMR is a technique to identify different types of hydrogen atoms that are present in the sample. ${{}^{1}}H$-NMR is a helpful method for knowing metabolic pathways in living cells. In this technique, the biological sample is dissolved in a solvent as the NMR reference that does not contain protons like deuterium oxide ($D_2O$) [see Fig. \ref{reff}(a)] \cite{legchenko2002review,jung20111h}.

The ${{}^{1}}H$-NMR spectrum shows some peaks from 0 to 14 ppm. 
Analyzing the NMR spectrum gives information about the number, positions, relative intensity, and splitting of signals \cite{levitt2013spin}. 
This information presents the number of protons in the sample and the number of hydrogens that generate the peaks. Moreover, they can show the amount of signal interaction with neighboring hydrogen atoms, which is known as spin splitting \cite{rugar2004single,westbrook2018mri}. 

The $n+1$ rule determines the splitting of proton signals that $n$ is the number of protons in the nearby nuclei. For example, when no hydrogen atom is around the nuclei, the splitting signal from the $n+1$ rule is a singlet. We present various splitting types in Fig. \ref{splitting}. According to Table \ref{nat1}, ${{}^{1}}H$-NMR is much more sensitive than ${{}^{13}}C$-NMR because of its highest natural abundance ($100\%$) and largest  $\gamma = 2.6752 \times 10^8 T^{-1}s^{-1}$ \cite{levitt1982broadband,slichter2013principles,krynicki1966proton}. 

\subsection{Carbon Nuclear Magnetic Resonance Spectroscopy}

NMR is also useful for carbon-based chemical samples like all living systems and organic compounds. Only about $1.11\% $ of all-natural carbons have $1/2$ quantum spin number with $\gamma=6.7283\times10^7T^{-1}s^{-1}$, which creates ${{}^{13}}C$ nucleus active in NMR imaging and makes it less sensitive than ${{}^{1}}H$-NMR, as is given in Table \ref{nat1}. The ${{}^{13}}C$-NMR sample should be mixed with a solvent as a reference compound, containing carbon such as TMS ($Si(CH_3)_4$) that makes protons entirely shielded [see Fig. \ref{reff}(b)] \cite{levitt2013spin,darbeau2006nuclear,grant1964carbon,schaefer1976carbon}. Compared to ${{}^{1}}H$-NMR spectrum, ${{}^{13}}C$-NMR peaks are spread from 0 to 200 ppm, which assists in the detection of distinct peaks with a low possibility of signal overlapping in NMR spectroscopy \cite{soni2021brain,graham1985anaerobic,sonnewald1993metabolism}.

\subsection{Phosphorus Nuclear Magnetic Resonance Spectroscopy}

${{}^{31}}P$ nucleus has $100\%$ natural abundance with $\gamma=1.0841\times10^8T^{-1}s^{-1}$, which behaves like ${{}^{1}}H$-NMR. However, ${{}^{31}}P$ chemical shifts cover a broader range compared to ${{}^{1}}H$-NMR. To achieve ${{}^{31}}P$-NMR signal, 
the NMR sample should be mixed with a solvent that contains phosphorus, such as phosphoric acid ($H_3PO_4$) [see Fig. \ref{reff}(c)]. 
From Table \ref{nat1}, we see that ${{}^{31}}P$-NMR has less sensitivity compared to ${{}^{1}}H$-NMR, and its peak is sharper than the peak of ${{}^{13}}C$-NMR. 

The high natural abundance and large $\gamma$ make ${{}^{31}}P$-NMR ideal for studying various systems such as phospholipid liposomes \cite{dubinnyi2006modeling}, exploring different food characteristics in food science like milk, green tea, and other foods \cite{gall2003nmr,renou1995nmr}. It is also helpful in detecting cancer tumors in  \textit{in vivo} models like breast cancer \cite{daly1987phospholipid} and lung cancer \cite{ng198231p}. 

\subsection{Nitrogen Nuclear Magnetic Resonance Spectroscopy}

${{}^{15}}N$-NMR is a powerful tool for providing useful information about samples that contain nitrogen atoms like RNA/DNA nucleobases \cite{barnwal2017applications}, molecular structure of organic compounds \cite{von198615n}, and soil organic matter \cite{mathers2000recent}. From Table \ref{nat1}, we see that ${{}^{15}}N$-NMR is much less sensitive compared to ${{}^{1}}H$-NMR and ${{}^{13}}C$-NMR, due to the  low natural abundance ($37\%$) of ${{}^{15}}N$ nucleus and its small $\gamma$ value ($\gamma=-2.713\times10^{7}T^{-1}s^{-1}$). Moreover, ${{}^{15}}N$-NMR chemical shift is significantly wide compared to ${{}^{13}}C$- and ${{}^{1}}H$-NMR, ranging from 0 to 900 ppm. In NMR studies with ${{}^{15}}N$,
Nitromethane ($CH_3NO_2$) is used as a standard external reference, as demonstrated in  Fig. \ref{reff}(e) \cite{khin1979some}.

\subsection{Fluorine Nuclear Magnetic Resonance Spectroscopy}

${{}^{19}}F$-NMR is a valuable tool for analyzing several amino acids, nucleotides, and sugars that contain fluorine. The ${{}^{19}}F$ nucleus has about $100\%$ natural abundance with large $\gamma=2.517\times10^{8}T^{-1}s^{-1}$, which has high sensitivity and broader chemical shift dispersion (ranging from -300 to 400 ppm) compared to ${{}^{1}}H$-NMR. Therefore, ${{}^{19}}F$-NMR is a convenient method to investigate important compounds of fluorine-containing pharmaceuticals \cite{okaru2017application,yu2013new}. In ${{}^{19}}F$-NMR,  Trichlorofluoromethane ($CFCl_3$) is usually used as a standard reference. The structure of $CFCl_3$ is shown in Fig. \ref{reff}(d).

\section{PHYSICS OF Magnetic Resonance Imaging}

\begin{figure}
\centering
	\includegraphics[width=\linewidth]{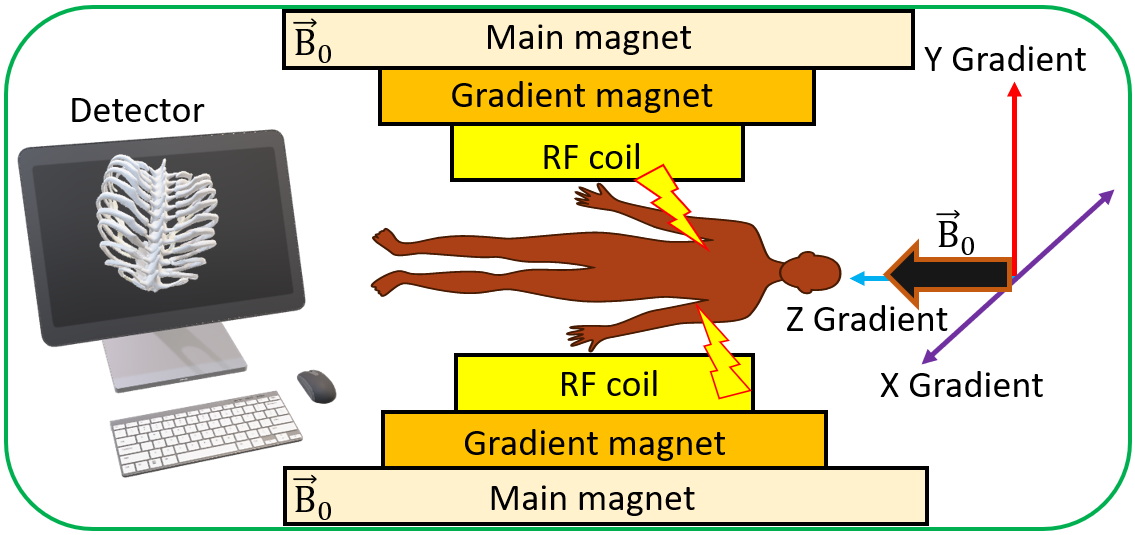}
	\caption{Scheme of MRI: The locations of main magnet $\vec{B_0}$ along the $z$-axis, gradient coils in the $x$-, $y$- and $z$-axis and RF receiver coils are shown in the Scheme. The MRI system is connected to a computer for processing the image of the body
	}
	\label{mri}
\end{figure}

MRI is based on the NMR technique utilized for medical imaging. Using MRI as a diagnostic method, we can learn about molecular fields and detect high-resolution images of soft tissues like the brain and cancer tumors \cite{plewes2012physics,bernstein2004handbook}. The human body comprises $70\%$ water, containing hydrogen and oxygen atoms. The hydrogen nuclei are the most abundant atoms in the human body. Therefore, we can visualize nuclei of atoms inside the body using MRI and generate powerful signals \cite{hirsch2015brute,filler2009history}.

The most significant advantage of MRI is its high signal-to-noise ratio because of the strong magnetic field utilized in this technology. MR imaging relies on the main magnetic field, a constant number usually as strong as 1.5-3T. According to the Zeeman effect, hydrogen atoms can align parallel or anti-parallel to $\vec{B_0}$. Regarding the Boltzmann distribution, more nuclei stay in the lower energy state than the higher energy state, which results in net magnetization [see Fig. \ref{mri}] \cite{canet1975time}.

MRI also contains gradient magnets that produce a magnetic field weaker than $\vec{B_0}$, known as the second part of the MRI machine. The gradient magnets generate a magnetic field over $\vec{B_0}$ in the $x$-, $y$- and $z$-axis. Moreover, the gradient magnets modify the strength of $\vec{B_0}$ and improve precession frequencies on the slice-selection gradient. This process generates spatial encoding for MRI signals \cite{pruessmann1999sense,friston1994analysis}. 

The spatial encoding can be obtained by projecting the signal from all slice spins along the gradient axes. Considering the spatial information obtained from this process, the axial images can be created by the data from the Z-gradient that runs along the long axis. The coronal images can be produced by the data from the Y-gradient, which is along the vertical axis. Finally, the sagittal images can be generated from X-gradient along the horizontal axis  \cite{plewes2012physics,friston1994analysis,okaru2017application}. 

The MRI machine also contains RF coils, which transmit RF pulses into the body and receive signals from the body to produce an image and record it on the screen. The RF coils are designed for specific body regions. These coils are designed to improve the signal-to-noise ratio, which in turn helps to generate better diagnostic images. 
In this scenario, an RF pulse, $\vec{B_1}$, that is orthogonal to $\vec{B_0}$ and has the precession frequency $\omega(x,y,x)= \gamma B_0 (x,y,x)$ should be applied, which turns the net magnetization toward the $x-y$ plane, as was explained in details for the NMR signal.

\section{Spin Echo Pulse sequence}

\begin{figure}
\centering
	\includegraphics[width=\linewidth]{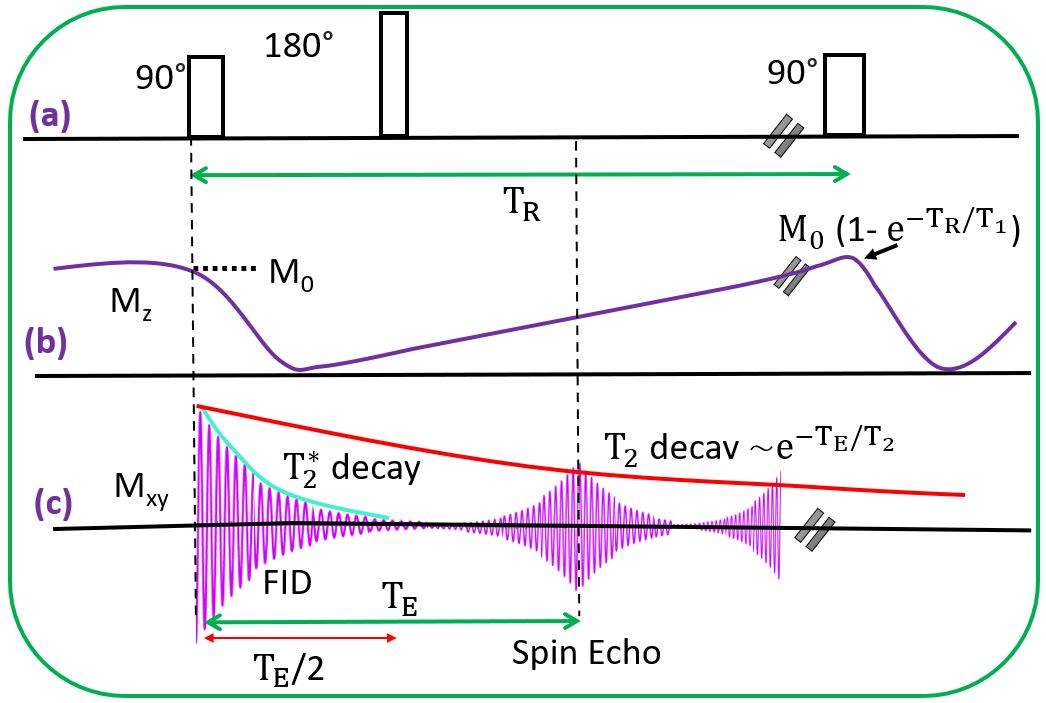}
	\caption{Scheme of a pulse sequence for producing the spin-echo signal: (a) represents 90-degree RF pulses and 180-degree pulse at ($T_E/2$) point. (b) The $M_z$ reduces from the start point, $M_0$, and grows along the $z$-axis until the next 90-degree pules (at $T_R$ point). (c) The $M_{xy}$ flips back into the $x-y$ plane, dephases, and records $T_2^{*}$ decays or FID signal. The spin echo signal measures and amplitude depend on $T_2$ at $T_E$ point. The spin echo repeats after $T_R$ point 
	}
	\label{Fig:13}
\end{figure}

The spin-echo sequence is an important pulse sequence used to create an image in MRI. The MRI machine can produce a high-quality image by repeating the pulse sequence several times, known as the spin-echo pulse sequence. The spin-echo pulse sequence refers to a 90-degree RF pulse followed by one or more 180-degree pulses to refocus the spins \cite{yao2009application,mosher2004cartilage}. This process is shown in Fig. \ref{Fig:13}. A spin-echo pulse sequence has two important parameters; the time between two 90-degree RF pulses is called repetition time or $T_R$, and the time between a 90-degree RF pulse and echo-formed signal in MRI, where an echo signal occurs is known as the time to echo or $T_E$. These are shown in Fig. \ref{Fig:13}(a)-(c). The magnitude of $M_z$ at $t=T_R$ is expressed as
\begin{equation}
M_{z}(T_R)= M_{0}\left(1-\exp\left(-\frac{T_R}{T_{1}}\right)\right).
\end{equation}
The FID signal reverses and recovers by applying a 180-degree pulse at $T_E/2$. Then, the magnetization relaxes along the $-z$-axis and returns to the $+z$-axis until it reaches its original value $M_0$. If a 90-degree RF pulse is applied, at any point during this process, $M_z$ rotates into $M_{xy}$ plane, then a signal can be observed \cite{basser1994estimation,antalek2002using}.  

The dephasing of magnetization causes loss of transverse net magnetization, decaying by $T_2$. If magnetization dephases further after the echo, the vectors refocus and form a second echo. By repeating this process, we can form multiple spin echoes.  After 90 degree RF pulse, according to Eq. (\ref{exp}), $M_{xy}$ at $T_E$  is
\begin{equation}
M_{x y}(T_E)=M_{x y}(0) \exp({-T_E / T_{2}}).
\end{equation}
The combination of $T_1$ and $T_2$ for a biological tissue can define signal output, S(r), that reads
\begin{equation} 
S(r) \simeq \rho \left(1-\exp\left(-\frac{T_R}{T_{1}}\right)\right)\exp\left(-\frac{T_E}{T_{2}}\right),
\end{equation}
where $\rho$ represents the density of protons per unit tissue. When the larger density of protons (hydrogen atoms) is measured, the brighter signal can be detected on the screen. Therefore, the higher proton density can enhance the image contrast and improve the final MRI signal \cite{styner2000parametric,chang1992technique,vovk2007review}.

The other concept that is related to the transverse relaxation is $T_2^*$ decay, which occurs between a 90-degree pulse and $T_E/2$ [see Fig. \ref{Fig:13}(c)]. In inhomogeneity of field determined by $\Delta B(r)$, which is time-independent, the spin system precesses with Larmor frequency $\omega(r)= \gamma \Delta B (r)$, and $M_{xy}(r,t)$ decays by 
\begin{equation}
 M_{x y}(r,t)=M_{x y}(r,0) \exp({-i\gamma \Delta B (r)t}).   
\end{equation}

The $T_2^*$ describes decay in the gradient echo (GRE) technique, significantly reducing the scan time. The relation between $T_2$ and $T_2^*$ is expressed by 
\begin{equation} 
\frac{1}{T_2^*} = \frac{1}{T_2} + \frac{1}{T_{2i}},
\end{equation}
in which $T_2$ corresponds to signal decays due to spin-spin relaxation, $T_{2i}$ is transverse relaxation due to static magnetic field inhomogeneities that lead to loss of coherence. The $T_2^*$ decays is always less than or equal to $T_2$ decays.

\section{Low field NMR and MRI properties}

The quest for high-resolution MRI has led researchers to perform novel high magnetic field techniques. A high magnetic field MRI is essential to tackle the inherent problem of low magnetic moments in nuclear spins. However, low-field MRI has also been shown to be possible due to progress in MRI techniques. One of the principal advantages of low-field MRI is that it provides a low-cost alternative to high-field instruments for MR imaging.

Moreover, the frequency of operation is much lower for low-field systems. The 0.06T system operates at approximately 2.46 MHz, while a 3T system operates closer to 128 MHz. Moreover, the magnetic field provided at the lower field is significantly homogeneous, making it an excellent clinical research resource \cite{griffiths2018introduction,albert1999t1,maxwell1864ii}. 

One shortcoming of low-filed MRI is that it is sensitive to noise originating from the environment, known as external electromagnetic interference (EMI) signals. There are several developments to prevent and remove EMI during the scan time. Another part of low-filed MRI is increasing relaxation time using a contrast agent \cite{feynman2011feynman,bardeen1957theory}. 

We note that some developments in MRI improve the SNR value at low-field MRI. Hyperpolarized (HP) MRI, where dynamic nuclear polarization (DNP) is used to increase the polarization of the spins in an MRI sample, compared with the thermal equilibrium state. HP MRI is better performed at low fields, as the spins relax back to their equilibrium state more slowly.

\section{Insights from Heisenberg uncertainty principle}

In NMR and MRI, the energy operator involved in the spectroscopy is known as energy transition, and the energy operator is called the Hamiltonian ($H$), which one can attain its eigenvectors and eigenvalues to analyze the system. For presenting the energy values for a nucleus, we should examine the wave function as a sum of all the different states aligned with or against the field, with some probability distributions, depending on the system characteristics \cite{abragam1983principles,schrodinger1926undulatory}. 

The Heisenberg time-energy uncertainty principle can provide the width of the NMR spectrum or the spectral width (SW) \cite{heisenberg1989encounters,rabi1938new,sakurai1995modern,khashami2024fundamentals}. The uncertainty principle for the nucleus in frequency $\Delta \nu$ and energy dispersion $\Delta E$ is given by 
\begin{equation} 
\Delta t \Delta E \approx  \Delta t (h \Delta \nu) \approx h. 
\end{equation} 

Also, the nucleus stays on the state at an uncertain lifetime $\Delta t$.  
Then, the peak width line (in Hz) for the nucleus, which is relaxed back to the lower state in time constant ${T_2}$ at frequency $\nu$, is rewritten as 
\begin{equation} 
\Delta \nu\approx \frac{1}{\Delta t} \approx \frac{1}{T_2}. 
\end{equation} 

We can detect the sharper peaks by recording the longer lifetime $\Delta t$ or increasing relaxation time ${T_2}$, which reduces the resonance linewidth $\Delta \nu$. $\Delta \nu$ refers to the range of the frequencies affected by the lifetime $\Delta t$.

\section{NMR and MRI application in medical science}

\begin{figure}
\centering
	\includegraphics[width=\linewidth]{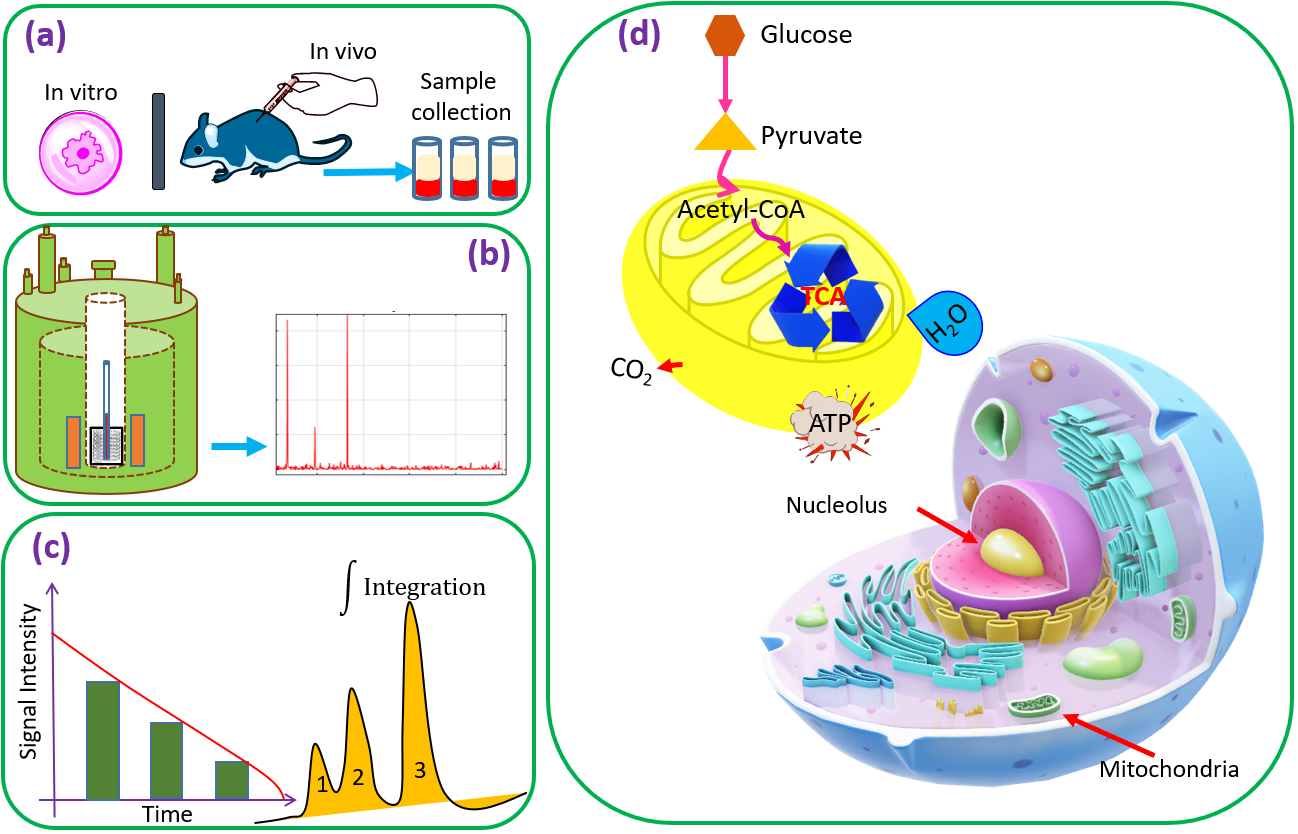}
	\caption{The basic of NMR data collection: (a) Sample preparation, (b) NMR signal detection, (c) Data analysis, (d) Metabolic pathway
	}
	\label{fig:1}
\end{figure}

The NMR or MRI sample is collected from either cell culture (\textit{in vitro}) or animal and human model (\textit{in vivo}) experiments, as we present in Fig. \ref{fig:1}(a). In \textit{in vitro} setup, the isotope molecules such as ${}^{1}\text{H}$, ${}^{13}\text{C}$, ${{{}^{31}}\text{P}}$, ${{{}^{15}}\text{N}}$, and ${{{}^{19}}\text{F}}$ are dissolved into special cell culture media and added to cell culture dishes to track their metabolism in the cell. As we mentioned earlier, all isotope molecules have specific standard external references, which are utilized to normalize the output signal from the scanner. 

After doping the powder in the cell culture dishes, the extracted cell sample is transferred into the NMR tube and placed in the magnet. Therefore, the final NMR spectrum is processed and displayed on a monitor for further metabolic analysis, as shown in Fig. \ref{fig:1}(b). 
Moreover, for \textit{in vivo} study,  the sample is prepared and injected into the body's soft tissue for further analysis.  

The NMR technique aims to precisely detect and quantify the metabolic changes in the cells or organs. For MR imaging, depending on the type of study performed, the sample is prepared, and an MRI scanner images the subject. Then, the MRI files are saved in a digital imaging and communications in medicine (DICOM) format and considered for further analysis,  as shown in Fig. \ref{fig:1}(c). 

To be precise about the data analysis steps, we must understand the cells' metabolic pathways and general physiology roles. As represented in Fig. \ref{fig:1}(d), glucose metabolism is most organs' primary energy source. Initially, complex glucose molecules are transported into cell membranes via specific transporters. Inside the cell, enzymes break down glucose into simpler forms, generating adenosine triphosphate (ATP), carbon dioxide ($CO_2$), and water ($H_2O)$ in the mitochondria. The mitochondria play a crucial role in this metabolic pathway. As a last product of the glycolysis pathway, the pyruvate, derived from glycolysis, travels through the TCA or citric acid cycle, contributing to the oxidative phosphorylation process that generates ATP.

\section{Conclusion}

Nuclear magnetic resonance (NMR) and magnetic resonance imaging (MRI) are versatile tools with broad applications from physics and chemistry to geology and medical studies. In this mini-review, we considered the concepts of NMR and MRI technologies from their fundamental origins to applications in medical science.  We started from a quantum mechanical basis and considered the significant importance of NMR and MRI in clinical research. Furthermore, we briefly introduced different types of NMR systems. We also investigated some of the most important applications of MRI techniques to provide valuable methods for visualizing the inside of the body and soft tissues. 

\bibliography{library}

\end{document}